%% file: dm-ddd-v_theta.tex
\documentclass[12pt, fleqn, a4paper]{article} % A4 size
\usepackage{hyperref}
\hypersetup{pdfstartview = FitH,
            pdfborder    = {0 0 0.5},
            colorlinks   = false
}
 \usepackage{epsfig, subcaption}
\usepackage{color}
\usepackage{clshan-math}
\def \lsim {\:\raisebox{-0.7 ex}{$\stackrel{\textstyle <}{\sim}$}\:}
\input{def-newcommands}
\begin{document}
\thispagestyle{empty}
\begin{flushright}
 March 2022
\end{flushright}
\begin{center}
{\Large\bf
 Incident Velocity--Recoil Angle Distribution and \\
 Angular Recoil--Energy Spectrum of               \\ \vspace{0.2 cm}
 3-Dimensional WIMP--Nucleus Scattering Events}   \\
\vspace*{0.7 cm}
 {\sc Chung-Lin Shan}                             \\
\vspace{0.5 cm}
 {\small\it
  Preparatory Office of
  the Supporting Center for
  Taiwan Independent Researchers                       \\ \vspace{0.05 cm}
  P.O.BOX 21 National Yang Ming Chiao Tung University,
  Hsinchu City 30099, Taiwan, R.O.C.}                  \\~\\~\\
 {\it E-mail:} {\tt clshan@tir.tw}
\end{center}
\vspace{2 cm}
\begin{abstract}

 In this paper,
 as a supplementary of
 our study on
 the angular distribution of
 the recoil flux of
 WIMP--scattered target nuclei
 and on
 that of
 the WIMP effective scattering velocity distribution,
 we investigate
 the scattering probability distribution of
 the WIMP incident velocity versus the nuclear recoil angle
 in narrow recoil energy windows
 for different WIMP masses and target nuclei.
 Our simulation results show that,
 not only the velocity distribution of
 incident halo WIMPs,
 but also a factor of the recoil angle
 could affect the scattering probability distribution of
 the available incident velocity--recoil angle combination
 in a given recoil energy window.
 Consequently,
 the 1-D WIMP ``effective'' velocity distribution
 corresponding to the considered narrow energy window
 would not be consistent with that
 cut simply from the (generating) velocity distribution of
 incident halo WIMPs.
 And its contribution
 to the differential WIMP--nucleus scattering event rate
 in the considered energy window
 could thus not be simply estimated by integrating over
 the 1-D theoretical velocity distribution (of
 entire halo WIMPs).

\end{abstract}
\clearpage
\section{Introduction}

 Direct Dark Matter (DM) detection experiments
 aiming to observe scattering signals of
 Weakly Interacting Massive Particles (WIMPs)
 off target nuclei
 would still be the most reliable experimental strategy
 for identifying Galactic DM particles
 and determining their properties
 \cite{SUSYDM96, Schumann19, Baudis20, Cooley21}.
 While most direct DM detection experiments
 measure only recoil energies
 deposited in underground detectors,
 the ``directional'' direct detection experiments
 could provide additional 3-dimensional information
 (recoil tracks and/or head--tail senses)
 of (elastic) WIMP--nucleus scattering events
 as a promising experimental strategy
 for discriminating WIMP signals
 from isotropic backgrounds
 and/or some incoming--direction--known astronomical events
 \cite{Ahlen09, Mayet16, Vahsen21}.

 As the preparation
 for our future study
 on the development of data analysis procedures
 for using and/or combining 3-D information
 offered by directional Dark Matter detection experiments,
 we have developed
 the double Monte Carlo scattering--by--scattering simulation package
 for the 3-D elastic WIMP--nucleus scattering process
 \cite{DMDDD-3D-WIMP-N}
 and studied
 the angular distributions of
 the recoil flux and energy of
 WIMP--scattered target nuclei
 in different celestial coordinate systems
 \cite{DMDDD-NR, DMDDD-NR-TAUP2021}.
 We have also introduced and demonstrated
 the target and WIMP--mass dependent ``effective'' velocity distribution of
 halo WIMPs
 (not only impinging on but also) scattering off target nuclei
 (in different celestial coordinate systems)
 \cite{DMDDD-fv_eff},
 which could be pretty different from
 the velocity distribution of
 incident Galactic WIMPs
 \cite{DMDDD-N, DMDDD-P}.

 During these works,
 the dependence of the recoil energy
 on the recoil angle
 and the incoming velocity of the scattering WIMPs
 as well as
 the dependence of the differential scattering cross section
 on the recoil angle
 and the nuclear form factor(s)
 have been noticed
 and discussed in detail in Ref.~\cite{DMDDD-InC}.
 It has been found that,
 firstly,
 for a given WIMP incident velocity,
 different recoil angles
 corresponding to different recoil energies
 should have different scattering probabilities
 and would contribute to the differential event rate differently
 \cite{DMDDD-InC}.
 Secondly,
 for a given recoil energy (window),
 different available incident velocity--recoil angle combinations
 should have different scattering probabilities
 and would contribute to the differential event rate differently
 \cite{DMDDD-InC}.
 These observations indicate further
 possible incompleteness of
 the conventionally used expressions for
 the (double) differential WIMP--nucleus scattering event rates
 in (directional) direct DM detection physics
 \cite{DMDDD-InC}
 as well as
 predict totally different
 3-D distribution patterns of
 the WIMP--induced nuclear recoil flux (and energy)
 from those provided in several earlier works
 (see e.g.~Refs.~\cite{Billard09, OHare14, Mayet16, OHare17, Vahsen21}).
 Hence,
 in this paper,
 we investigate the scattering distribution of
 the WIMP incident velocity versus the nuclear recoil angle
 in narrow recoil energy windows
 for different target nuclei and WIMP masses,
 in order to provide more detailed information on
 its contribution to
 the (double) differential scattering event rates.

 The remainder of this paper is organized as follows.
 In Sec.~2,
 we review briefly
 some basic considerations to factors,
 which could affect the scattering probability of
 an incident velocity--recoil angle combination.
 Then
 the scattering distributions
 on the incident velocity vs.~recoil angle plane
 as well as
 the angular distributions of
 the recoil fluxes
 for a light and a heavy WIMP masses
 scattering off the light $\rmF$
 and the heavy $\rmXe$ nuclei
 will be presented and discussed in Secs.~3 and 4,
 respectively.
 We summarize our observations
 in Sec.~5.

%
% 1/3
 \input{sec-basics}
%
%

%
% 2/3
 \input{sec-mchi-0020}
%
%

%
% 3/3
 \input{sec-mchi-0200}
\section{Summary}

 In this paper,
 as a supplementary of
 our study on
 the angular distribution of
 the recoil flux of
 WIMP--scattered target nuclei
 and on
 that of
 the WIMP effective scattering velocity distribution,
 we investigated
 the scattering probability distribution of
 the WIMP incident velocity versus the nuclear recoil angle
 in narrow recoil energy windows
 for different WIMP masses and target nuclei.

 As argued in detail in Ref.~\cite{DMDDD-InC},
 our simulations show that,
 not only the velocity distribution of
 incident halo WIMPs,
 but also the factor of the recoil angle
 appearing in the differential scattering cross section
 could affect the scattering probability distribution of
 the available incident velocity--recoil angle combination
 constrained by the bounds of a (narrow) recoil energy window.
 As consequences,
 firstly,
 the scattering probability always reduces strongly
 (almost vanishes)
 around zero recoil angle (\mbox{$\eta \simeq 0$}).
 Secondly,
 the 1-D WIMP effective velocity distributions
 corresponding to different narrow energy windows
 would not be consistent with each other
 nor those cut simply
 from the generating velocity distribution of
 incident halo WIMPs
 by the minimal--required WIMP incident velocities.
 And their contributions
 (from the same velocity range)
 to the differential WIMP--nucleus scattering event rates
 in different narrow energy windows
 could not be the same
 nor simply estimated by integrating over
 the 1-D theoretical velocity distribution (of
 entire halo WIMPs).

 Moreover,
 once the WIMP mass is as light as only a few tens GeV,
 for both of light and heavy target nuclei,
 the most recoil angles
 would nevertheless be small enough
 to maintain the angular distributions of
 the induced recoil events
 pretty concentrated
 (around the opposite direction of
  the theoretical WIMP incident direction).
 In contrast,
 once the WIMP mass is as heavy as a few hundreds GeV,
 the most actually--induced recoil angles (of
 events observed in low energy windows)
 could be pretty large.
 This means that
 a large number of observed recoil events
 would be strongly deflected
 (almost perpendicularly)
 and
 the corresponding angular recoil--flux distribution patterns
 could then be pretty wide and flat,
 especially
 when heavy nucleus like I or Xe
 is used as our target.

 On the other hand,
 our simulation results
 demonstrated in the geocentric Galactic coordinate system
 show clearly
 the target and WIMP--mass dependence of
 (the variation of)
 the angular recoil--flux distributions
 in different narrow energy windows.
 Firstly,
 with the increasing WIMP mass,
 the angular recoil--flux distributions
 become flatter
 and the differences between
 the distribution patterns with different target nuclei
 become also larger.
 Secondly,
 while
 for a WIMP mass of a few tens GeV,
 (the decrease of) the recoil fluxes
 on the inner sky
 are clearly larger (and sharper)
 than those on the outer sky,
 once the WIMP mass is a few hundreds GeV,
 (the decrease of) the recoil fluxes
 on the outer sky
 could inversely be larger (and sharper)
 than those on the inner sky;
 the heavier the target nuclei,
 the higher the upper limit of the energy window,
 under which
 one can observe this ``reverse'' inner--outer--asymmetry.
 These indicate a possibility of
 (analytic and perhaps model--independent) reconstruction of
 the WIMP mass
 by comparing and/or combining
 the angular recoil--energy spectra
 with different target nuclei.

 In summary,
 in this work
 we studied
 the elastic WIMP--nucleus scattering process
 and the angular recoil--energy spectrum
 in more details.
 Hopefully,
 this work
 could help our colleagues
 to develop methods and analyses for reconstructing properties of
 Galactic Dark Matter particles
 by using 3-dimensional information
 offered by directional direct detection experiments.

\subsubsection*{Acknowledgments}

 This work
 was strongly encouraged by
 the ``{\it Researchers working on
 e.g.~exploring the Universe or landing on the Moon
 should not stay here but go abroad.}'' speech.

%
%
% References
%
 \input{sec-references}
%

%
%
%
\end{document}

%% file: def-newcommands.tex
\newcommand{\mchi}        {m_{\chi}}
\newcommand{\mN}          {m_{\rm N}}
\newcommand{\mrN}         {m_{\rm r, N}}
\newcommand{\vmin}        {v_{\rm min}}
\newcommand{\sigmaSI}     {\sigma_0^{\rm SI}}
\newcommand{\sigmaSD}     {\sigma_0^{\rm SD}}
\newcommand{\sigmapSI}    {\sigma_{\chi {\rm p}}^{\rm SI}}
\newcommand{\armp}        {a_{\rm p}}
\newcommand{\armn}        {a_{\rm n}}
\newcommand{\FSIQ}        {F_{\rm SI}^2(Q)}
\newcommand{\FSDQ}        {F_{\rm SD}^2(Q)}
\newcommand{\VchiLab}     {{\bf v}_{\chi, {\rm Lab}}}
\newcommand{\vchiLab}     {     v _{\chi, {\rm Lab}}}
\newcommand{\Echi}        {     E _{\chi}}
\newcommand{\thetaNRchi}  {\theta_{\rm N_R, \chi_{in}}}
 \newcommand{\rmF}  {\rmXA{F}  {19}}
 \newcommand{\rmAr} {\rmXA{Ar} {40}}
 \newcommand{\rmGe} {\rmXA{Ge} {73}}
 \newcommand{\rmXe} {\rmXA{Xe}{129}}
 \newcommand{\rmW}  {\rmXA{W} {183}}
\newcommand{\OnlineSpectrumvtheta} [3] {
\href{http://www.tir.tw/phys/hep/dm/amidas-2d/amidas-2d.php%
      ?amidas_2D_function=v_theta%
      &mode_animation=spectrum2D%
      &target=\Target%
      &mchi=#1%
      &period=periodA%
      &Q_range=#2%
      &event_No=500}
     {#3}
}
\newcommand{\OnlineSpectrumNRang} [2] {
\href{http://www.tir.tw/phys/hep/dm/amidas-2d/amidas-2d.php%
      ?amidas_2D_function=NR_ang%
      &mode_NR=NR_ang_2D\%2B%
      &frame=#2%
      &mode_animation=spectrum2D%
      &target=\Target%
      &mchi=#1%
      &period=periodA%
      &Q_range=0_5_10_15_20%
      &event_No=500}
     {\begin{minipage} {17.2 cm}
       \begin{center}
          \begin{subfigure} [c] {4.2 cm}
           \includegraphics [width = 4.2 cm] {NR_ang-\Target-\WIMPmass-#2-0000_0050-0500-00000}%
           \vspace{-0.15 cm}
          \caption{ 0 --  5 keV}
          \end{subfigure}
          \begin{subfigure} [c] {4.2 cm}
           \includegraphics [width = 4.2 cm] {NR_ang-\Target-\WIMPmass-#2-0050_0100-0500-00000}%
           \vspace{-0.15 cm}
          \caption{ 5 -- 10 keV}
          \end{subfigure}
          \begin{subfigure} [c] {4.2 cm}
           \includegraphics [width = 4.2 cm] {NR_ang-\Target-\WIMPmass-#2-0100_0150-0500-00000}%
           \vspace{-0.15 cm}
          \caption{10 -- 15 keV}
          \end{subfigure}
          \begin{subfigure} [c] {4.2 cm}
           \includegraphics [width = 4.2 cm] {NR_ang-\Target-\WIMPmass-#2-0150_0200-0500-00000}%
           \vspace{-0.15 cm}
          \caption{15 -- 20 keV}
        \end{subfigure}
       \end{center}
      \end{minipage}}
}
\newcommand{\InsertPlotvthetaN} [2] {
\begin{figure} [t!]
\begin{center}
 \OnlineSpectrumvtheta
  {#1}
  {0_5_20_100}
  {\begin{minipage} {15 cm}
    \begin{center}
       \begin{subfigure} [c] {6.3  cm}
        \includegraphics [width = 6.3  cm] {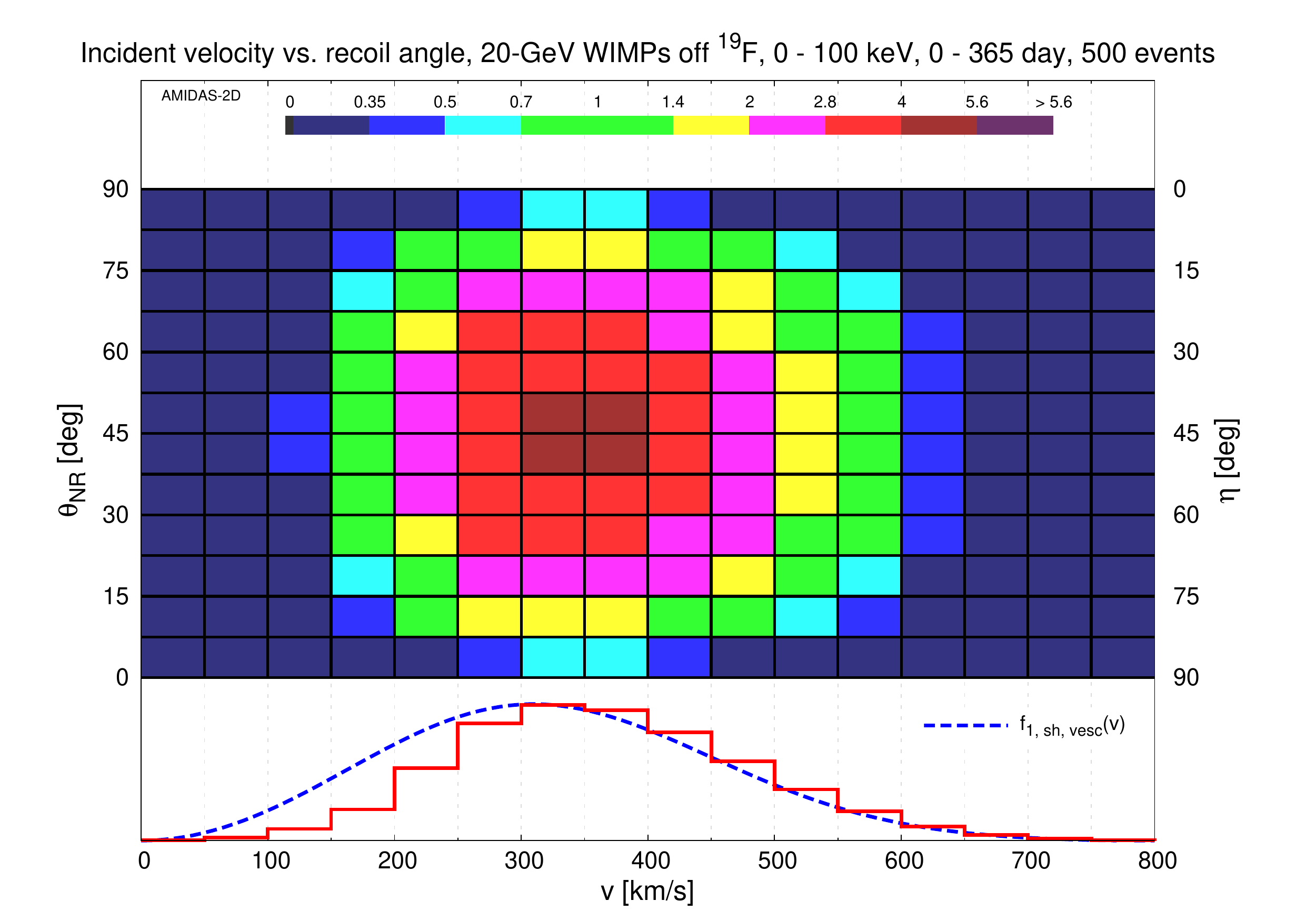}%
        \vspace{-0.15 cm}
       \caption{500 events in 0 -- 100 keV}
       \end{subfigure}
       \hspace{ 2.1  cm}
       \begin{subfigure} [c] {6.3  cm}
        \includegraphics [width = 6.3  cm] {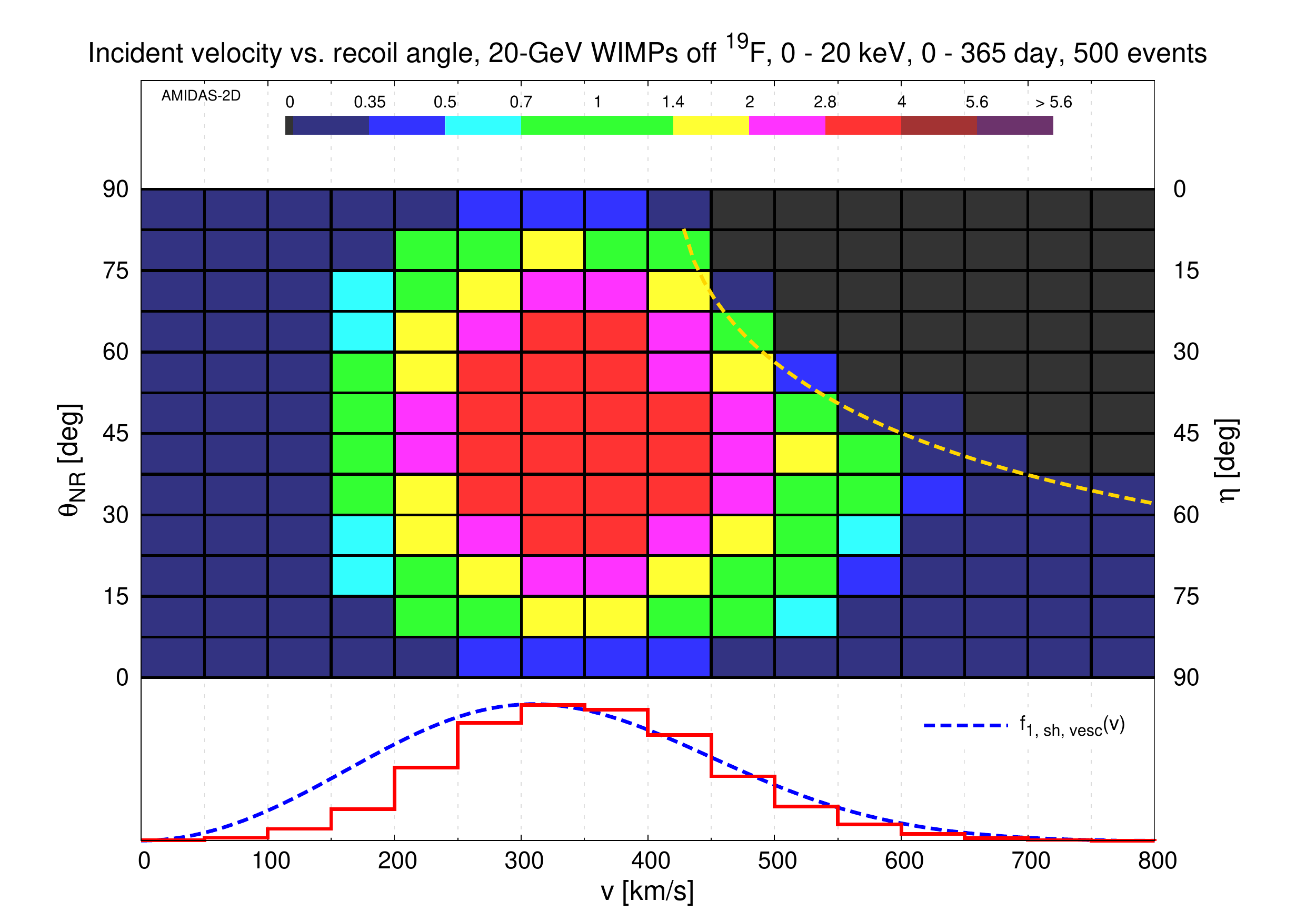}%
        \vspace{-0.15 cm}
       \caption{500 events in 0 -- 20 keV}
       \end{subfigure}
    \end{center}
   \end{minipage}}
 \\ \vspace{ 0.25 cm}
 \OnlineSpectrumvtheta
  {#1}
  {0_5_10_15_20}
  {\begin{minipage} {17.2 cm}
    \begin{center}
       \begin{subfigure} [c] {4.2 cm}
        \includegraphics [width = 4.2 cm] {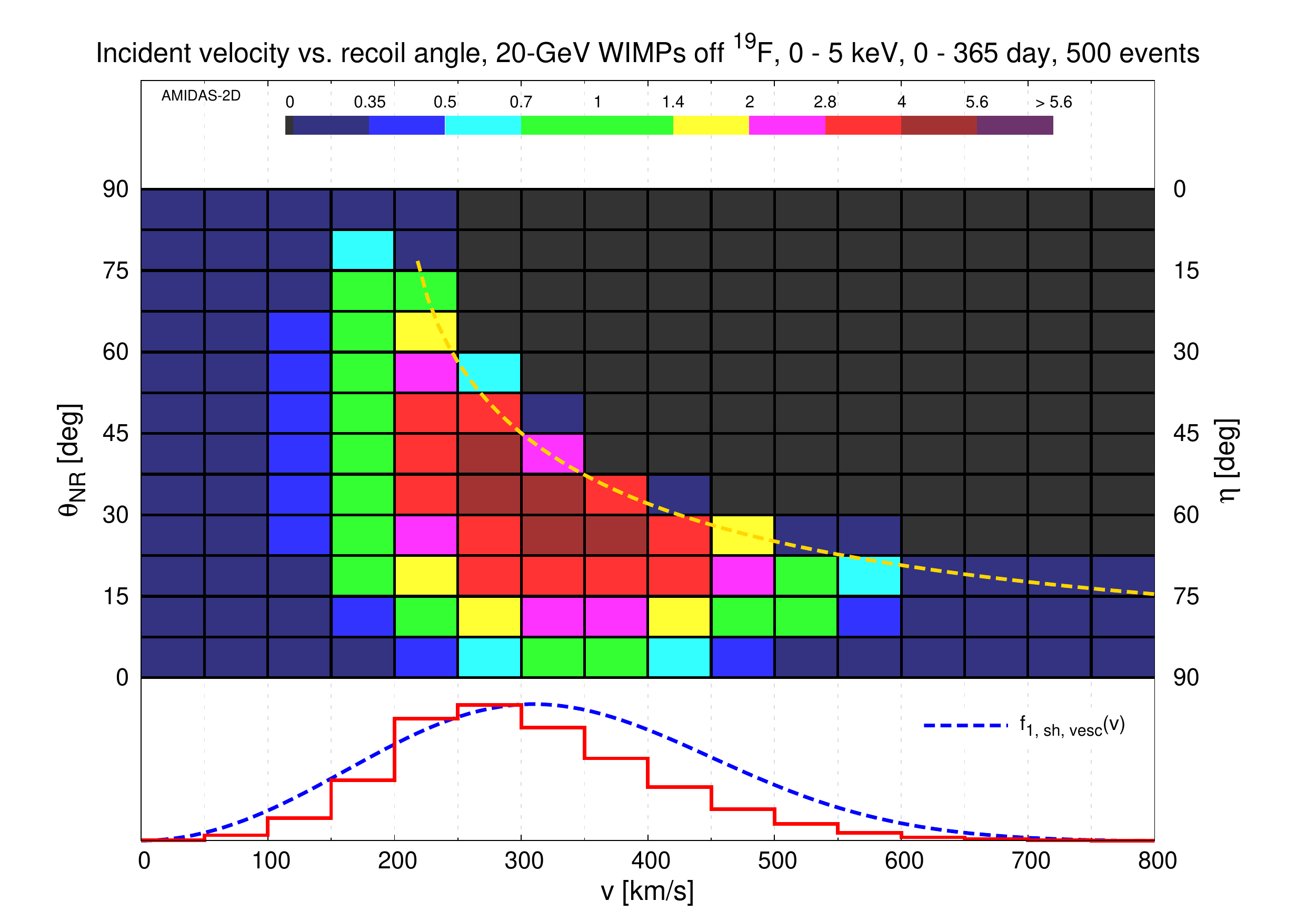}%
        \vspace{-0.15 cm}
       \caption{ 0 --  5 keV}
       \end{subfigure}
       \begin{subfigure} [c] {4.2 cm}
        \includegraphics [width = 4.2 cm] {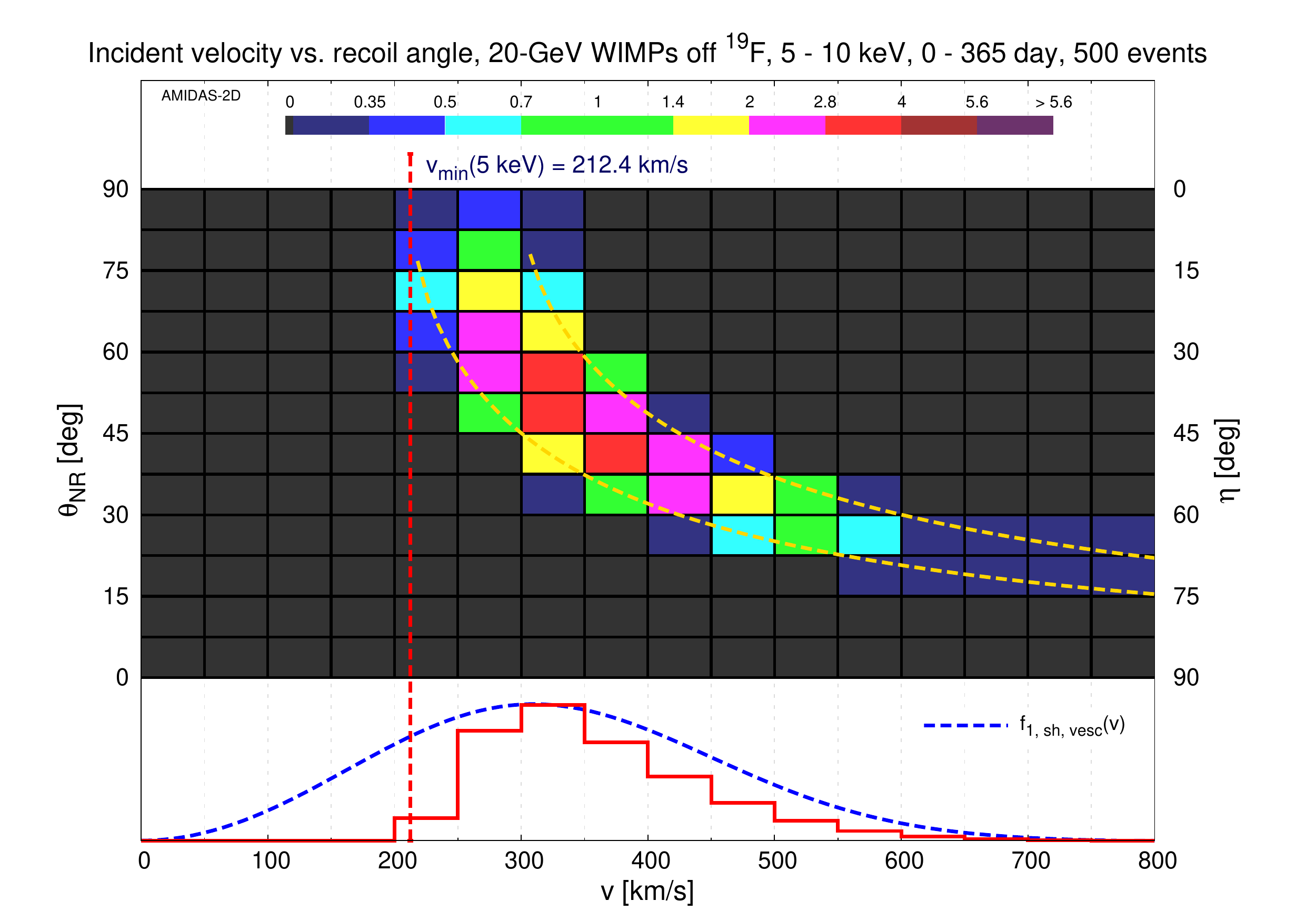}%
        \vspace{-0.15 cm}
       \caption{ 5 -- 10 keV}
       \end{subfigure}
       \begin{subfigure} [c] {4.2 cm}
        \includegraphics [width = 4.2 cm] {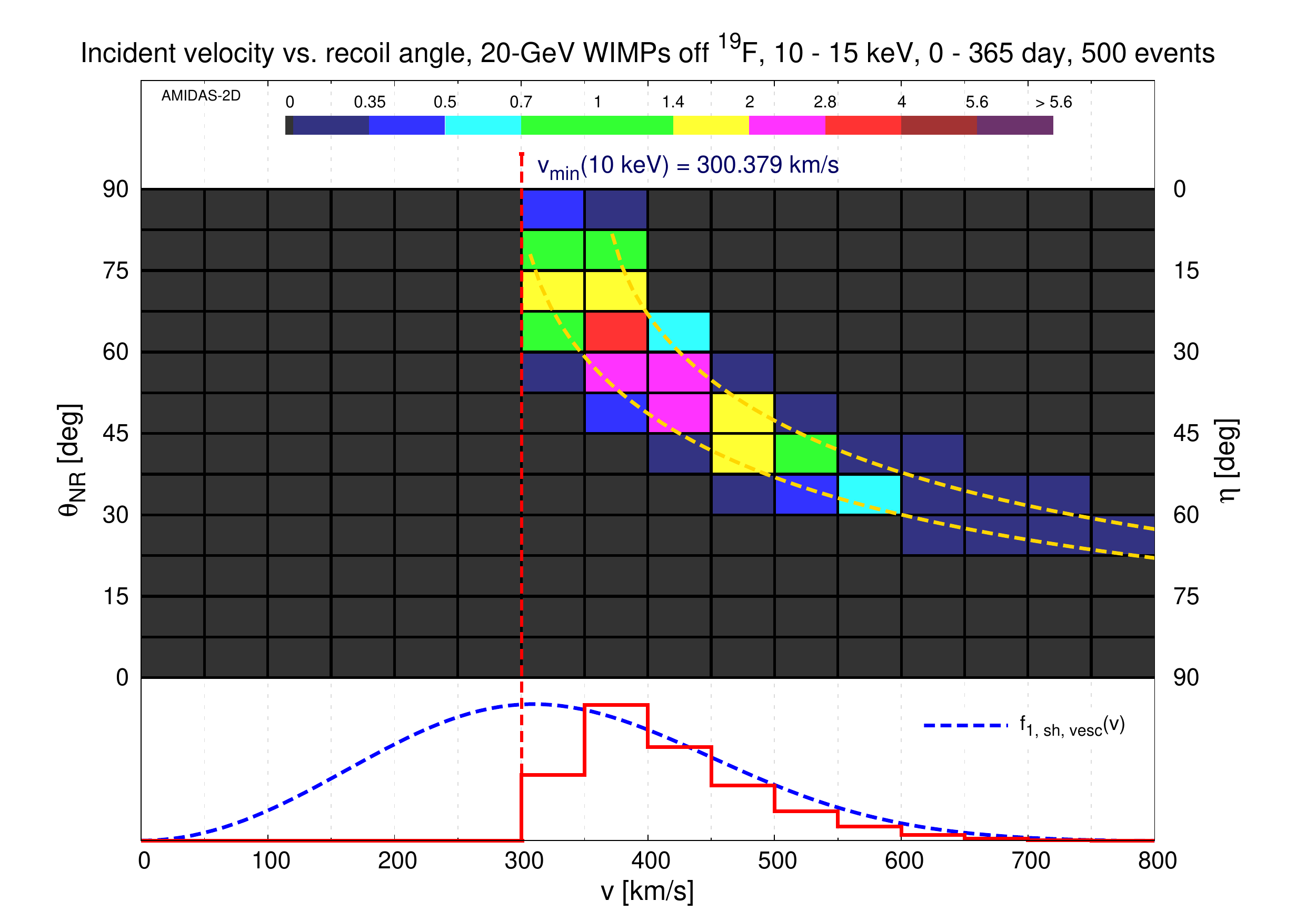}%
        \vspace{-0.15 cm}
       \caption{10 -- 15 keV}
       \end{subfigure}
       \begin{subfigure} [c] {4.2 cm}
        \includegraphics [width = 4.2 cm] {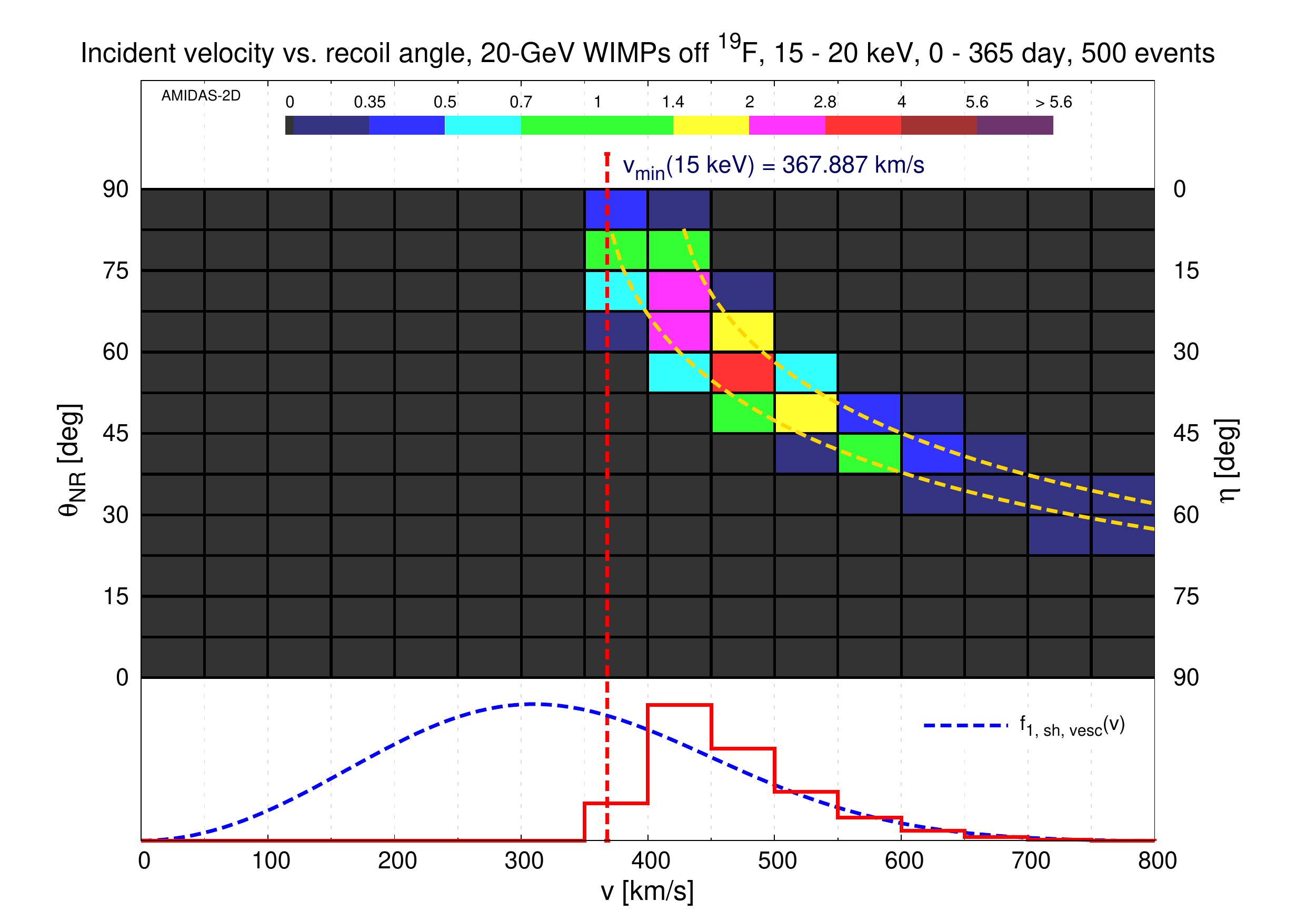}%
        \vspace{-0.15 cm}
       \caption{15 -- 20 keV}
       \end{subfigure}
    \end{center}
   \end{minipage}}
 \\ \vspace{ 0.25 cm}
 \OnlineSpectrumNRang
  {#1}
  {Eq}
 \\ \vspace{ 0.25 cm}
 \OnlineSpectrumNRang
  {#1}
  {geoG}
 \\ \vspace{-0.25 cm}
\end{center}
\caption{
 #2
}
\label{fig:v_theta-N-\Target-\WIMPmass-0500-00000}
\end{figure}
}

%% file: sec-basics.tex
%
%
\section{Basics}
\label{sec:basics}

 This work is based on
 our double Monte Carlo scattering--by--scattering simulations
 for 3-D elastic WIMP--nucleus scattering
 described in detail in Refs.~\cite{DMDDD-3D-WIMP-N, DMDDD-NR}.
 In this section,
 we review only briefly
 some basic considerations to factors,
 which could affect the scattering probability of
 an incident velocity--recoil angle combination.

 Consider microscopically
 one incident halo WIMP of a mass $\mchi$
 moving with an incoming velocity $\VchiLab$
 and scattering off a target nucleus of a mass $\mN$.
 The kinetic energy of
 the incident (and scattering) WIMP
 in the laboratory (detector at rest) coordinate system
 can be given by
\beq
     \Echi
  =  \frac{1}{2} \mchi |\VchiLab|^2
  =  \frac{1}{2} \mchi  \vchiLab ^2
\~.
\label{eqn:Echi}
\eeq
 Then
 the recoil energy of the scattered target nucleus
 (in the laboratory coordinate system)
 can be expressed as
 a function of
 the recoil angle $\eta$,
 or,
 equivalently,
 the ``equivalent'' recoil angle $\thetaNRchi$ by
 \cite{DMDDD-3D-WIMP-N}
\beqn
     Q
  =  \bbrac{\frac{4 \mchi \mN}{(\mchi + \mN)^2} \~ \cos^2(\eta)}
     \Echi
 \=  \bbrac{\afrac{2 \mrN^2}{\mN} \vchiLab^2}
     \cos^2(\eta)
     \non\\
 \=  \bbrac{\afrac{2 \mrN^2}{\mN} \vchiLab^2}
     \sin^2(\thetaNRchi)
\~.
 \label{eqn:QQ_eta}
 \label{eqn:QQ_thetaNRchi}
\eeqn
 Here
\(
         \mrN
 \equiv  \mchi \mN / \abrac{\mchi + \mN}
\)
 is the WIMP--nucleus reduced mass,
 $\eta$
 is the recoil angle of
 the scattered nucleus
 (the angle
  between the recoil direction
  and the WIMP incoming velocity)
 \cite{DMDDD-3D-WIMP-N},
 and
 $\thetaNRchi = \pi / 2 - \eta$
 (i.e.~the complementary angle of $\eta$)
 is the elevation of the nuclear recoil direction
 in the incoming--WIMP coordinate system
 \cite{DMDDD-3D-WIMP-N}.

\begin{figure} [t!]
\begin{center}
 \includegraphics [width = 12 cm] {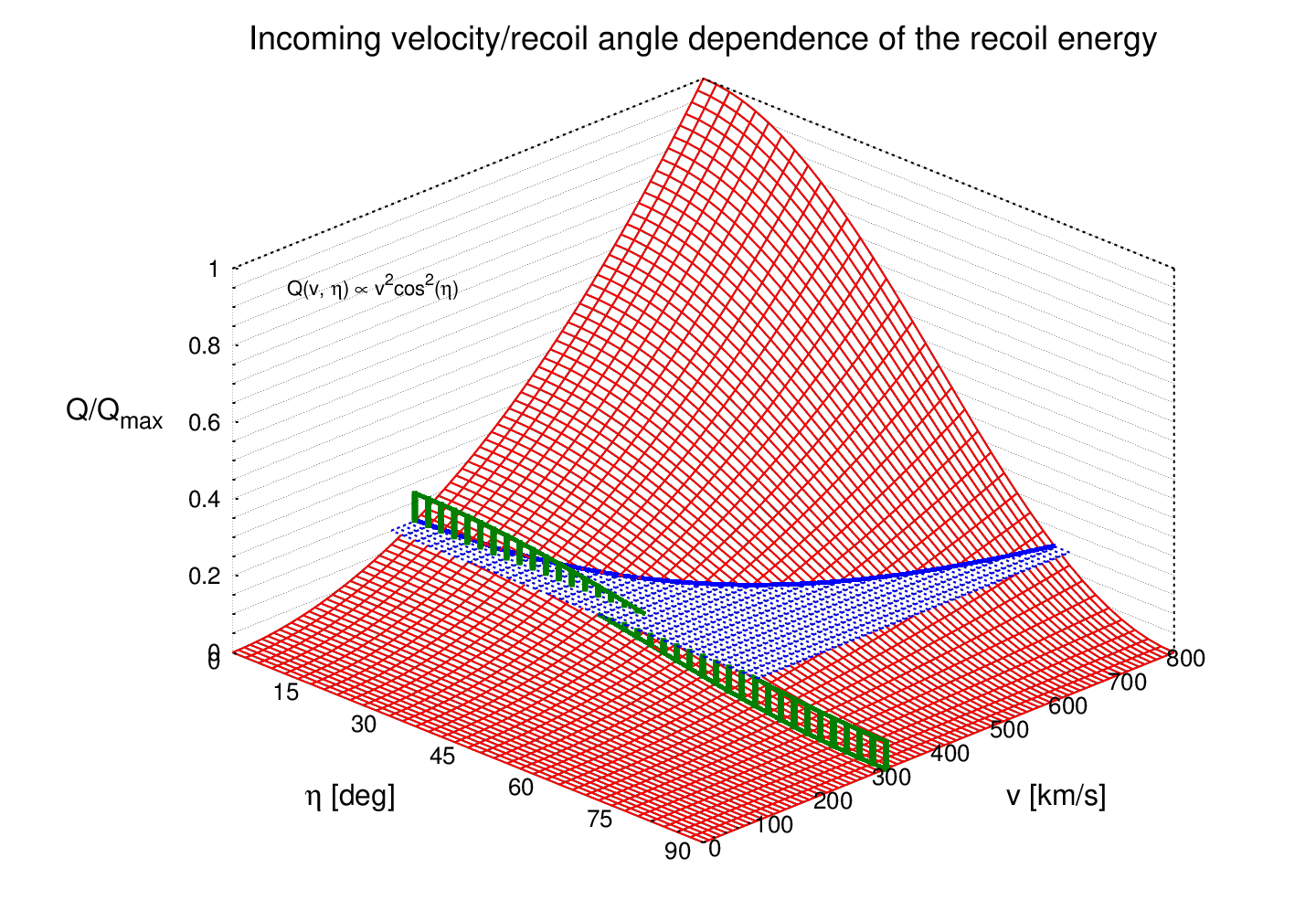}
\\
\vspace{-6.5 cm}
\begin{picture}(9.5 , 6.5 )
\color{blue}
\put(8.5 , 3.4 ){\makebox(0.6 , 0.4 ){$Q^{\ast}$}}
\put(1.7 , 3.7 ){\makebox(0.8 , 0.4 ){$\vmin^{\ast}$}}
\end{picture}
 \vspace{-0.25 cm}
\end{center}
\caption{
 The 2-D dependence of the recoil energy on
 the incoming velocity of incident WIMPs $\vchiLab$
 and the recoil angle $\eta$
 given in Eq.~(\ref{eqn:QQ_thetaNRchi}).
 The green vertical fence indicates
 the given WIMP incident velocity $\vmin^{\ast}$,
 while
 the blue plane is
 the equal--recoil--energy plane of
 the corresponding energy $Q^{\ast}$.
}
\label{fig:Q_eta-v_m-Q_m}
\end{figure}
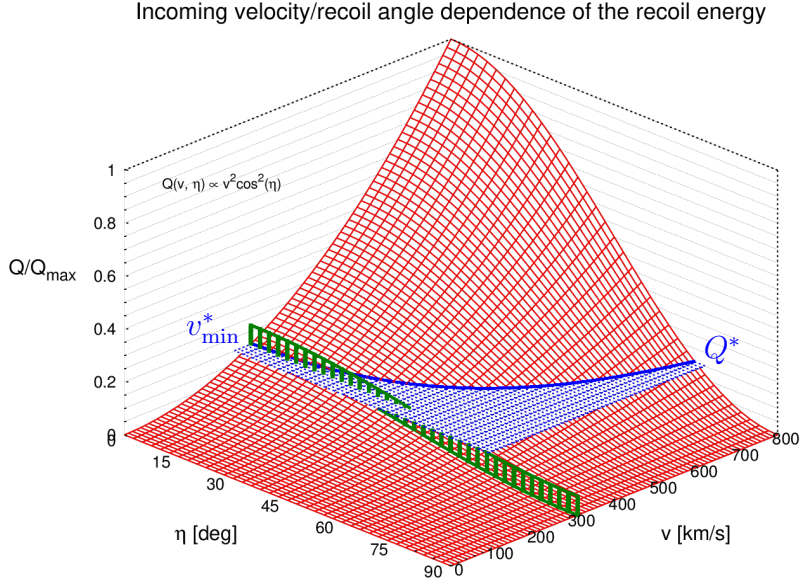

 Conventionally,
 people only use Eq.~(\ref{eqn:QQ_thetaNRchi})
 with the minimal (maximal equivalent) recoil angle of $\eta = 0$
 ($\thetaNRchi = 90^{\circ}$)
 to define the minimal--required incoming velocity
 of incident WIMPs,
 which can transfer the considered recoil energy $Q^{\ast}$
 to a target nucleus:
\beq
     \vmin^{\ast}
  =  \vmin(Q^{\ast})
  =  \sfrac{\mN}{2 \mrN^2} \~ \sqrt{Q^{\ast}}
  =  \alpha \sqrt{Q^{\ast}}
\~,
\label{eqn:vmin}
\eeq
 where
\(
         \alpha
\equiv  \sqrt{\mN / 2 \mrN^2}
\)
 is the transformation constant.
 However,
 as discussed in detail in Ref.~\cite{DMDDD-InC},
 the expression (\ref{eqn:QQ_thetaNRchi})
 indicates that
 the WIMP--induced nuclear recoil energy
 should be considered as a two--variable function of
 the WIMP incident velocity $\vchiLab$
 and the recoil angle $\eta$,
 as shown in Fig.~\ref{fig:Q_eta-v_m-Q_m}.
 For incident WIMPs
 with (monotonically) the given incoming velocity $\vmin^{\ast}$,
 $Q^{\ast}$ can be considered as
 the ``maximal transferable'' recoil energy.
 This implies,
 firstly,
 that,
 due to the nuclear form factor suppression,
 the probability of WIMP scattering events
 with a small or even zero (large equivalent) recoil angle
 and thus a large recoil energy around $Q^{\ast}$
 should be strongly reduced
 \cite{DMDDD-InC}.

 Secondly,
 along the (blue) equal--recoil--energy--$Q^{\ast}$ contour,
 one has
\cheqna
\beq
     \eta(Q^{\ast}, \vchiLab)
  =  \cos^{-1}\afrac{\alpha \sqrt{Q^{\ast}}}{\vchiLab}
\~,
\label{eqn:eta_vchiLab}
\eeq
 or,
 equivalently,
\cheqnb
\beq
     \thetaNRchi(Q^{\ast}, \vchiLab)
  =  \sin^{-1}\afrac{\alpha \sqrt{Q^{\ast}}}{\vchiLab}
\~.
\label{eqn:thetaNRchi_vchiLab}
\eeq
\cheqn
 Although
 the strength of the nuclear form factor suppression
 (depending only on the recoil energy $Q^{\ast}$)
 is equal,
 the velocity distribution of incident halo WIMPs
 affects the scattering probabilities of
 different $\vchiLab$--$\eta$ combinations%
\footnote{
 Note that,
 from Eqs.~(\ref{eqn:eta_vchiLab})
 and (\ref{eqn:thetaNRchi_vchiLab}),
 one can find
 ``maximal (minimal) available'' (equivalent) recoil angles
 for transferring the considered recoil energy $Q^{\ast}$
 as
\cheqna
\beq
     \eta_{\rm max}(Q^{\ast})
  =  \cos^{-1}\afrac{\alpha \sqrt{Q^{\ast}}}{v_{\chi, {\rm cutoff}}}
\~,
\label{eqn:eta_max}
\eeq
 and
\cheqnb
\beq
     \theta_{\rm N_R, \chi_{in}, min}(Q^{\ast})
  =  \sin^{-1}\afrac{\alpha \sqrt{Q^{\ast}}}{v_{\chi, {\rm cutoff}}}
\~,
\label{eqn:thetaNRchi_max}
\eeq
\cheqn
 where $v_{\chi, {\rm cutoff}} \simeq 800$ km/s
 is a cut--off velocity of incident halo WIMPs
 (in the Equatorial/laboratory coordinate systems).
}.

 Thirdly and most importantly,
 from the conventionally used expression for
 the differential WIMP--nucleus scattering cross section
 as the function of
 the absolute value of the momentum transfer
 from the incident WIMP to the scattered nucleus
 \cite{SUSYDM96, Schumann19, Baudis20}:
\beq
     d\sigma
  =  \frac{1}{\vchiLab^2}
     \afrac{1}{4 \mrN^2}
     \bbigg{\sigmaSI F_{\rm SI}^2(q) + \sigmaSD F_{\rm SD}^2(q)} dq^2
\~,
\label{eqn:dsigma_dq2}
\eeq
 with
\(
     q
  =  \sqrt{2 \mN Q}
\),
 the spin--independent/dependent (SI/SD) total cross sections
 ignoring the nuclear form factor suppressions,
 $\sigma_0^{\rm (SI, SD)}$,
 as well as
 the elastic nuclear form factors
 corresponding to the SI/SD WIMP interactions,
 $F_{\rm (SI, SD)}(q)$,
 the differential cross section $d\sigma$
 with respect to the differential (equivalent) recoil angles
 $d\eta$ ($d\thetaNRchi$)
 have been derived as
 \cite{DMDDD-3D-WIMP-N, DMDDD-InC}
\cheqna
\beq
     \vDd{\sigma}{\eta}
  =  \bbigg{\sigmaSI \FSIQ + \sigmaSD \FSDQ}
     \sin(2 \eta)
\~,
\label{eqn:dsigma_deta}
\eeq
 and
\cheqnb
\beq
     \Dd{\sigma}{\thetaNRchi}
  =  \bbigg{\sigmaSI \FSIQ + \sigmaSD \FSDQ}
     \sin(2 \thetaNRchi)
\~,
\label{eqn:dsigma_dthetaNRchi}
\eeq
\cheqn
 respectively.
 Note that
 we use here the absolute value of $d\sigma / d\eta$,
 since the recoil energy $Q$ decreases
 while the recoil angle $\eta$ increases.
 Remind also that
 the recoil energy $Q$ here
 should be considered as
 the function of $\vchiLab$ and $\eta$ ($\thetaNRchi$)
 given in Eq.~(\ref{eqn:QQ_eta}).

 Moreover,
 by taking into account
 the proportionality of
 the WIMP flux
 to the WIMP incident velocity,
 the scattering probability of
 incident halo WIMPs
 {\em moving with the given incoming velocity $\vchiLab$}
 and scattering off target nuclei
 going into an (equivalent) recoil angle between
 $\eta \pm d\eta / 2$
 ($\thetaNRchi \pm d\thetaNRchi / 2$)
 with a recoil energy of $Q \pm dQ / 2$
 can generally be expressed by
 \cite{DMDDD-3D-WIMP-N, DMDDD-InC}
\cheqna
\beq
     f_{\rm N_R}(\vchiLab, \eta)
  =  \frac{\vchiLab}{v_{\chi, {\rm cutoff}}}
     \bbigg{\sigmaSI \FSIQ + \sigmaSD \FSDQ}
     \sin(2 \eta)
\~,
\label{eqn:f_NR_eta}
\eeq
 and
\cheqnb
\beq
     f_{\rm N_R}(\vchiLab, \thetaNRchi)
  =  \frac{\vchiLab}{v_{\chi, {\rm cutoff}}}
     \bbigg{\sigmaSI \FSIQ + \sigmaSD \FSDQ}
     \sin(2 \thetaNRchi)
\~.
\label{eqn:f_NR_thetaNRchi}
\eeq
\cheqn
 These two expressions
 indicate clearly that,
 firstly,
 ``head--on'' (zero--recoil--angle, $\eta = 0$) scattering events
 should be impossible.
 Secondly,
 the scattering probability distribution
 on the incident velocity versus the (equivalent) recoil angle
 ($\vchiLab$--$\eta$ ($\thetaNRchi$)) plane
 would be (much) more complicated
 as people though earlier.
 Remind that
 so far it could be understood that
 the scattering probability distribution
 depends at least on
 the velocity distribution of incident WIMPs,
 the nuclear form factor suppression of target nuclei,
 which depends further on
 their atomic mass and the WIMP mass,
 as well as
 the recoil--angle constraint on
 the differential scattering cross section.

%% file: sec-mchi-0020.tex
%
%
\section{Scattering by light 20-GeV WIMPs}
\label{sec:mchi-0020}

 As discussed in detail in Ref.~\cite{DMDDD-InC}
 and reviewed briefly in Sec.~\ref{sec:basics},
 in different (narrow) recoil energy window,
 the scattering probability of
 each available WIMP incident velocity--nuclear recoil angle combination
 and its contribution to
 the differential WIMP--nucleus scattering event rate
 should be different.
 Hence,
 in this and the next sections,
 following our earlier work on
 the angular distribution of
 the WIMP--induced nuclear recoil flux
 \cite{DMDDD-NR}%
\footnote{
 Remind that
 we generate first a {\em 3-dimensional velocity} of
 each incident WIMP
 in the Galactic coordinate system
 according to the theoretical isotropic Maxwellian velocity distribution
 and
 transform it to the laboratory coordinate system
 \cite{DMDDD-3D-WIMP-N}.
 Then,
 in the laboratory
 (more precisely,
  the incoming--WIMP) coordinate system,
 we generate an {\em equivalent recoil angle} of
 a scattered target nucleus
 and validate this candidate scattering event
 according to the criterion (\ref{eqn:f_NR_thetaNRchi})
 \cite{DMDDD-3D-WIMP-N}.
 Remind also that,
 in our double Monte Carlo simulation procedure,
 the induced recoil energy is determined
 by the WIMP incident velocity $\vchiLab$
 and the equivalent recoil angle $\thetaNRchi$
 through Eq.~(\ref{eqn:QQ_thetaNRchi}).
},
 we reduce the simulated energy window
 to a width of only 5 keV
 and then investigate the scattering probability distribution
 on the WIMP incident velocity versus nuclear recoil angle plane.

 In this section,
 we consider at first
 the case of a light WIMP mass of $\mchi = 20$ GeV.
 Two spin--sensitive nuclei
 used frequently
 in (directional) direct DM detection experiments:
 $\rmF$ and $\rmXe$
 have been considered as our targets%
\footnote{
 Although Xe is (so far) not used
 in directional direct detection experiments,
 our simulation results
 shown in Secs.~\ref{sec:Xe129-0020} and \ref{sec:Xe129-0200}
 would be similar to those
 with the $\rmXA{I}{127}$ nucleus.
}.
 As in our earlier works
 presented in Refs.~%
 \cite{DMDDD-N, DMDDD-P, DMDDD-NR, DMDDD-fv_eff, DMDDD-NR-TAUP2021},
 the SI (scalar) WIMP--nucleon cross section
 in Eqs.~(\ref{eqn:f_NR_eta}) and (\ref{eqn:f_NR_thetaNRchi})
 has been fixed as \mbox{$\sigmapSI = 10^{-9}$ pb},
 while
 the effective SD (axial--vector) WIMP--proton/neutron couplings
 have been tuned as $\armp = 0.01$
 and $\armn = 0.7 \armp = 0.007$,
 respectively.
 5,000 experiments
 with 500 {\em accepted} events on average
 (Poisson--distributed)
 in one entire year
 in one experiment
 for one target nucleus
 have been simulated.

 For readers' reference,
 all simulation results demonstrated in this paper (and more)
 can be found ``in animation''
 on our online (interactive) demonstration webpage
 \cite{AMIDAS-2D-web}.

\subsection[20-GeV WIMPs off $\rmF$ nuclei]
           {\boldmath
            20-GeV WIMPs off $\rmF$ nuclei}
\label{sec:F19-0020}

 We consider at first
 the case of light WIMPs of $\mchi = 20$ GeV
 scattering off light $\rmF$ target nuclei.

 \def \Target       {F19}
 \def \WIMPmass     {0020}
 \InsertPlotvthetaN
  {20}
  {(a) -- (f)
   The scattering probability distributions of
   the available $\vchiLab$--$\thetaNRchi$ ($\eta$) combinations
   for the case of 20-GeV WIMPs scattering off $\rmF$
   in six different recoil energy windows.
   500 accepted
   WIMP scattering events on average
   (Poisson--distributed)
   in one entire year
   in one experiment
   for one target nucleus
   have been simulated
   and binned into 16 $\vchiLab$ bins
   $\times$ 12 angular $\thetaNRchi$ ($\eta$) bins.
   While
   the dashed golden curve(s) in the upper part of each plot indicate(s)
   the equal--recoil--energy contour(s) of
   (the lower and) the upper bound(s) of
   the considered energy window,
   the dashed red vertical line indicates
   $\vmin$
   estimated by Eq.~(\ref{eqn:vmin})
   with the lower bound of
   the energy window.
   The dashed blue curve in the lower part of each plot indicates
   the generating
   shifted Maxwellian velocity distribution of halo WIMPs
   given in Eq.~(7) of Ref.~\cite{DMDDD-fv_eff},
   whereas
   the solid red histogram is
   the actual (effective) velocity distribution of
   the scattering WIMPs.
   Note that
   each histogram has been normalized
   to have the same height as the theoretical curve.
   (g) -- (j) and (k) -- (n)
   The corresponding angular distributions of
   the WIMP--induced nuclear recoil flux
   (in unit of the all--sky average value)
   observed in the Equatorial and the ``geocentric'' Galactic coordinate systems,
   respectively
   \cite{DMDDD-N, DMDDD-3D-WIMP-N}.
   The magenta diamonds (dark--green stars) indicate
   the (opposite) direction of the Solar Galactic movement
   \cite{DMDDD-N, DMDDD-3D-WIMP-N}.%
   \vspace{-1.05 cm}
   }

 In Figs.~\ref{fig:v_theta-N-F19-0020-0500-00000}(a)
 and \ref{fig:v_theta-N-F19-0020-0500-00000}(b),
 we show
 the scattering probability distributions of
 the available combinations of
 the incident WIMP velocity $\vchiLab$ versus
 the (equivalent) recoil angle $\thetaNRchi$ (left axis) and also $\eta$ (right axis)
 in the recoil energy windows
 between 0 and 100 (20) keV,
 respectively%
\footnote{
 Interested readers can click each row of the plots
 in Figs.~\ref{fig:v_theta-N-F19-0020-0500-00000},
 \ref{fig:v_theta-N-Xe129-0020-0500-00000},
\ref{fig:v_theta-N-F19-0200-0500-00000},
 and \ref{fig:v_theta-N-Xe129-0200-0500-00000}
 to open the corresponding webpage of
 animated demonstrations
 (for more considered target nuclei/WIMP masses).
}.
 The horizontal color bar on the top of the plot
 indicates
 the mean value of the recorded event number
 (averaged over all simulated experiments)
 in each $\vchiLab$--$\thetaNRchi$ ($\eta$) bin
 in unit of the average value of the {\em non--empty} bins
 (500 events / 192 (156) bins
  $\cong$ \mbox{2.60 (3.21) events/bin},
  respectively).
 The dashed golden curve
 in Fig.~\ref{fig:v_theta-N-F19-0020-0500-00000}(b)
 indicates
 the equal--recoil--energy contour of
 $\eta(Q = 20~{\rm keV}, \vchiLab)$
 given in Eq.~(\ref{eqn:eta_vchiLab})
 (i.e.,~the blue interaction curve in Fig.~\ref{fig:Q_eta-v_m-Q_m}).
 As references,
 in the lower part of each plot,
 the dashed blue curve indicates
 the theoretically predicted
 shifted Maxwellian velocity distribution
 given in Eq.~(7) of Ref.~\cite{DMDDD-fv_eff}
 for generating incident halo WIMPs,
 whereas
 the solid red histogram is
 the actual (effective) velocity distribution of
 the scattering WIMPs,
 which has been normalized
 to have the same height as the theoretical curve.

 It can be found that,
 due to the factor of $\sin(2 \eta)$
 in the expression (\ref{eqn:f_NR_eta})
 for the scattering probability
 as well as
 the Gaussian--like WIMP velocity distribution,
 the most frequent incident velocity--recoil angle combinations
 would be around $\eta \simeq 45^{\circ}$
 and $\vchiLab \simeq 340$ km/s,
 which is a little bit larger than
 the average velocity of entire halo WIMPs of
 $\bar{v}_{\rm Lab} \simeq 331$ km/s,
 as can also be seen directly by
 comparing the (red) histograms
 with the (blue) theoretical curves.

 Moreover,
 in Figs.~\ref{fig:v_theta-N-F19-0020-0500-00000}(c)
 to \ref{fig:v_theta-N-F19-0020-0500-00000}(f),
 we slice the energy window up to 5 keV each%
\footnote{
 Note that
 500 accepted events on average
 in ``each (5-keV)'' energy window
 have been simulated.
}.
 Now
 the dependence of the scattering probability distributions
 on the available incident velocity--recoil angle combinations
 can be seen more clearly.
 Here,
 as a reference,
 the dashed red vertical line in each plot indicates
 the minimal--required WIMP incident velocity
 $\vmin$
 estimated by Eq.~(\ref{eqn:vmin})
 with the lower bound of
 the energy window.
 As expected,
 all available incident velocity--recoil angle combinations
 are in the area
 enclosed by the equal--recoil--energy contour(s).
 Constrained by the (most important) $\sin(2 \eta)$ factor,
 the scattering probabilities always reduce strongly
 (almost vanish)
 around zero recoil angle ($\eta \simeq 0$),
 although
 the distributions of the recoil angle
 shift towards smaller $\eta$
 with the raising energy window.
 Additionally,
 the (red) WIMP ``effective'' velocity distributions
 corresponding to different energy windows
 seem also to be different
 from each other
 and those cut simply
 from the (blue) compared shifted Maxwellian velocity distribution
 by the minimal--required WIMP incident velocities
 (the dashed red vertical lines).
 This can be observed more clearly
 in Figs.~\ref{fig:v_theta-N-F19-0020-0500-00000}(e)
 and \ref{fig:v_theta-N-F19-0020-0500-00000}(f)
 (as well as
  in Figs.~\ref{fig:v_theta-N-Xe129-0020-0500-00000}(d)
  and \ref{fig:v_theta-N-Xe129-0020-0500-00000}(e)).

 Correspondingly,
 Figs.~\ref{fig:v_theta-N-F19-0020-0500-00000}(g)
 to \ref{fig:v_theta-N-F19-0020-0500-00000}(j)
 show
 the angular distributions of
 the WIMP--induced nuclear recoil flux
 (in unit of the all--sky average value)
 observed in the Equatorial coordinate system
 \cite{DMDDD-N, DMDDD-3D-WIMP-N}.
 The dark--green stars indicate
 the opposite direction of the Solar Galactic movement
 \cite{DMDDD-N, DMDDD-3D-WIMP-N}:
 42.00$^{\circ}$S, 50.70$^{\circ}$W.
 The differences
 between the recoil--flux distribution patterns
 can be seen obviously.
 The higher the considered energy window,
 the smaller the (most) frequent recoil angle $\eta$
 and thus
 the more concentrated the nuclear recoil flux.
 It would however be interesting to notice that
 the maxima of
 the recoil flux
 in all four considered energy windows
 would {\em not} match
 the so--called ``WIMP--wind'' direction
 (the opposite direction of the Solar Galactic movement),
 but shift {\em northerly}.

 Finally,
 for readers' comparisons with the plots
 shown in e.g.~Refs.~\cite{Billard09, 
                           OHare14,
                           Mayet16,
                           Vahsen21},
 in Figs.~\ref{fig:v_theta-N-F19-0020-0500-00000}(k)
 to \ref{fig:v_theta-N-F19-0020-0500-00000}(n),
 we provide
 the corresponding angular recoil--flux distributions
 (in unit of the all--sky average value)
 observed in a ``geocentric'' Galactic coordinate system%
\footnote{
 It is basically the conventional astronomical Galactic coordinate system
 \cite{DMDDD-N}:
 its coordinate axes are parallel to
 our Galactic coordinate system,
 but its origin is located
 at the Earth's center.
 Hence,
 the transformation between our Equatorial and geocentric Galactic coordinate systems
 is a pure rotation
 and the relative velocity of the Earth in the Galaxy
 is discarded.
}.
 While
 the dark--green stars indicate
 the opposite direction of the Solar Galactic movement
 \cite{DMDDD-N, DMDDD-3D-WIMP-N}:
 0.60$^{\circ}$S, 98.78$^{\circ}$W,
 the magenta diamonds indicate additionally
 the moving direction of the Solar system:
 0.60$^{\circ}$N, 81.22$^{\circ}$E.
 Besides
 a clear north--south--sky symmetry,
 the recoil--flux distribution patterns
 on the east and west side of the sky
 divided by the opposite direction of the Solar Galactic movement
 (the green star)
 show asymmetric intenses and decreasing gradients:
 (the decrease of) the recoil fluxes
 on the inner (east) sky
 are clearly larger (and sharper)
 than those on the outer (west) sky.

\subsection[20-GeV WIMPs off $\rmXe$ nuclei]
           {\boldmath
            20-GeV WIMPs off $\rmXe$ nuclei}
\label{sec:Xe129-0020}

 In this subsection,
 we consider a heavy nucleus $\rmXe$
 as our detector target.

\begin{figure} [t!]
\begin{center}
 \includegraphics [width = 9.5 cm] {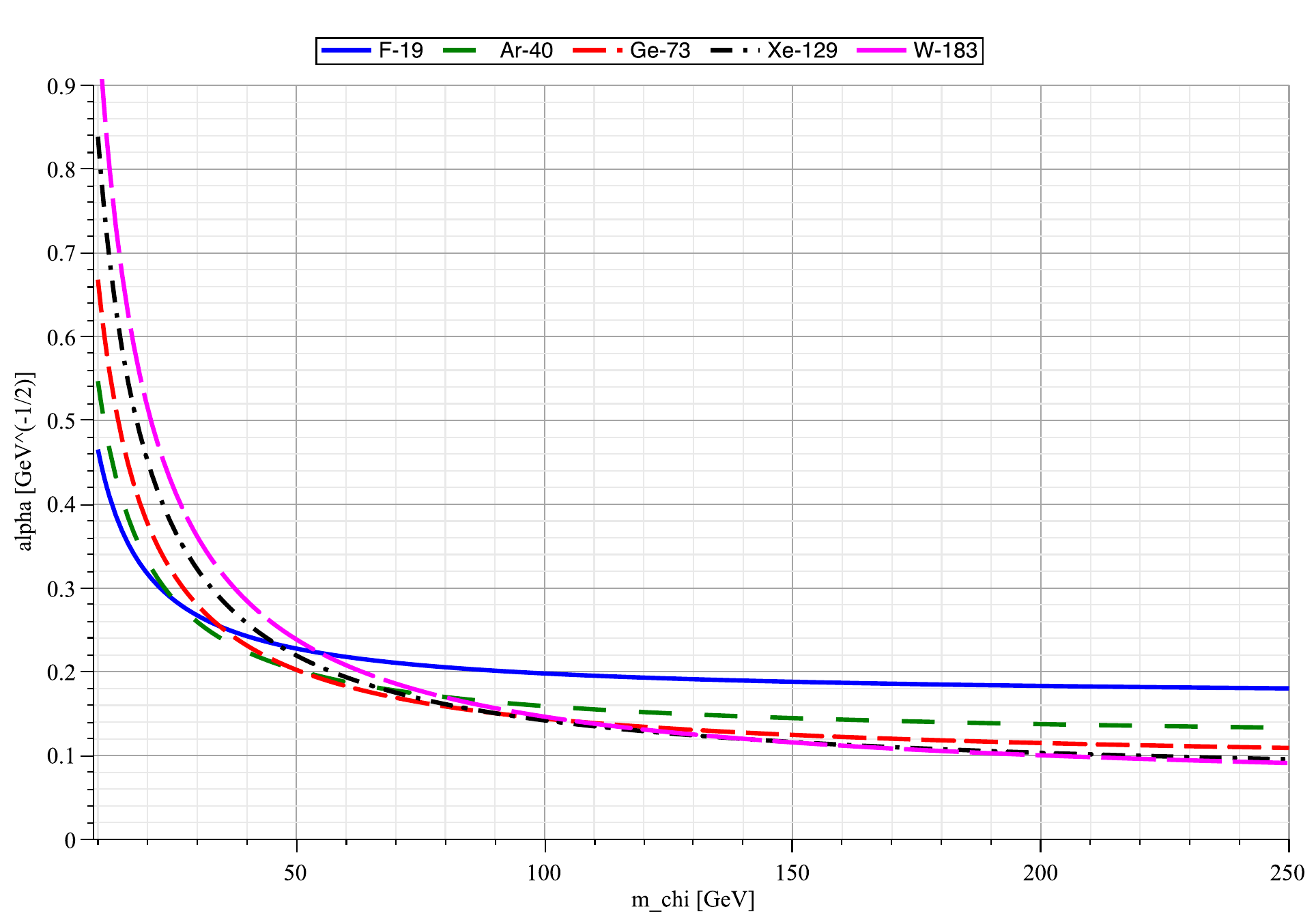}
 \vspace{-0.5 cm}
\end{center}
\caption{
 The WIMP--mass dependence of
 the transformation constant $\alpha$
 given in Eq.~(\ref{eqn:vmin}).
 Five frequently used target nuclei:
 $\rmF$     (solid        blue),
 $\rmAr$    (rare--dashed green),
 $\rmGe$    (dashed       red),
 $\rmXe$    (dash--dotted black),
 and $\rmW$ (long--dashed magenta)
 have been considered.
}
\label{fig:alpha-mchi-F-Ar-Ge-Xe-W}
\end{figure}
 \def \Target       {Xe129}
 \InsertPlotvthetaN
  {20}
  {As Figs.~\ref{fig:v_theta-N-F19-0020-0500-00000}:
   the mass of incident halo WIMPs is $\mchi = 20$ GeV,
   except that
   a heavy nucleus $\rmXe$
   has been considered as our target.%
   }

 As shown in Fig.~\ref{fig:alpha-mchi-F-Ar-Ge-Xe-W},
 for WIMP masses $\mchi \lsim 45$ GeV,
 the transformation constant $\alpha$
 given in Eq.~(\ref{eqn:vmin})
 for the $\rmF$ nucleus (solid blue)
 is smaller than that for the $\rmXe$ nucleus (dash--dotted black).
 Thus
 the boundaries of
 the available $\vchiLab$--$\eta$ combinations
 shown in Figs.~\ref{fig:v_theta-N-Xe129-0020-0500-00000}(b)
 to \ref{fig:v_theta-N-Xe129-0020-0500-00000}(f)
 shift now towards higher $\vchiLab$ and smaller $\eta$ area.
 This implies that,
 as shown in Figs.~\ref{fig:v_theta-N-Xe129-0020-0500-00000}(g)
 to \ref{fig:v_theta-N-Xe129-0020-0500-00000}(j),
 once the WIMP mass is $\lsim 45$ GeV,
 the angular recoil--flux distribution of
 the scattered xenon nuclei
 should be more concentrated than
 those with fluorine.
 The northern shifts would also become smaller.
 In Figs.~\ref{fig:v_theta-N-Xe129-0020-0500-00000}(k)
 to \ref{fig:v_theta-N-Xe129-0020-0500-00000}(n)),
 while
 the concentration of
 the recoil--flux distributions
 with the raising recoil energy window
 as well as
 the north--south--sky symmetry
 and the inner--outer--sky asymmetry
 can be observed more clearly,
 the decrease of the recoil fluxes
 on the inner and the outer skies
 are also sharper
 with the xenon nuclei
 than with fluorine.

 Moreover,
 as mentioned in Sec.~\ref{sec:F19-0020},
 the (red) WIMP effective velocity distributions
 corresponding to different energy windows
 (especially
  those shown in Figs.~\ref{fig:v_theta-N-Xe129-0020-0500-00000}(d)
  and \ref{fig:v_theta-N-Xe129-0020-0500-00000}(e))
 would obviously differ
 from those cut simply
 from the (blue) compared generating velocity distribution.
 This would confirm clearly
 our theoretical prediction
 discussed in detail in Ref.~\cite{DMDDD-InC} that
 the 1-D effective velocity distribution of
 the {\em scattering} WIMPs
 would not be identical to
 the velocity distribution of
 incident halo WIMPs
 and thus its contributions
 (from the same velocity range)
 to the differential WIMP--nucleus scattering event rates
 in different energy ranges
 could not be simply estimated by integrating over
 the 1-D theoretical velocity distribution (of
 entire halo WIMPs).

%% file: sec-mchi-0200.tex
%
%
\section{Scattering by heavy 200-GeV WIMPs}
\label{sec:mchi-0200}

 In this section,
 we raise the mass of incident WIMPs to $\mchi = 200$ GeV.

\subsection[200-GeV WIMPs off $\rmF$ nuclei]
           {\boldmath
            200-GeV WIMPs off $\rmF$ nuclei}
\label{sec:F19-0200}

 As in Sec.~\ref{sec:F19-0020},
 we consider at first
 the light nucleus $\rmF$
 as our detector target.

 \def \Target       {F19}
 \def \WIMPmass     {0200}
 \InsertPlotvthetaN
  {200}
  {As Figs.~\ref{fig:v_theta-N-F19-0020-0500-00000}:
   the light nucleus $\rmF$
   has been considered as our target,
   except that
   the mass of incident halo WIMPs is raised to $\mchi = 200$ GeV.%
   }

 While
 the difference
 between
 the $\vchiLab$--$\eta$ distributions
 and the (red) effective velocity distributions
 in Figs.~\ref{fig:v_theta-N-F19-0020-0500-00000}(a)
 and \ref{fig:v_theta-N-F19-0200-0500-00000}(a)
 for the full energy range between 0 and 100 keV
 is pretty small,
 except of the four bins
 around $\vchiLab \simeq 350$ km/s
 and $\eta$ $\simeq 45^{\circ}$,
 with a much lower upper cut--off of the analyzed energy window of 20 keV,
 the difference
 between Figs.~\ref{fig:v_theta-N-F19-0020-0500-00000}(b)
 and \ref{fig:v_theta-N-F19-0200-0500-00000}(b)
 can be observed clearly.
 A shift of
 (the average velocity of)
 the effective velocity distribution of
 the scattering WIMPs
 towards a lower velocity
 and,
 consequently,
 a shift to
 a {\em larger} most frequent recoil angle of
 $\eta \simeq 52.5^{\circ}$
 can be seen directly.

 More precisely,
 Figs.~\ref{fig:v_theta-N-F19-0200-0500-00000}(c)
 to \ref{fig:v_theta-N-F19-0200-0500-00000}(f)
 show that,
 with the raising WIMP mass
 and thus the reducing transformation constant $\alpha$,
 the most frequent recoil angle ($\eta$)
 would be pretty large,
 especially
 in the low energy range of $Q \le 10$ keV.
 Consequently,
 the corresponding angular recoil--flux distributions
 in Figs.~\ref{fig:v_theta-N-F19-0200-0500-00000}(g)
 to \ref{fig:v_theta-N-F19-0200-0500-00000}(j)
 as well as
 those
 in Figs.~\ref{fig:v_theta-N-F19-0200-0500-00000}(k)
 to \ref{fig:v_theta-N-F19-0200-0500-00000}(n)
 show pretty widely spread patterns
 (wider than those in Figs.~\ref{fig:v_theta-N-F19-0020-0500-00000}).

 It should be important to notice here that,
 with (light target nuclei like) $\rmF$,
 the variation of
 the angular recoil--flux distributions
 in four considered narrow energy windows
 shown in Figs.~\ref{fig:v_theta-N-F19-0200-0500-00000}(k)
 to \ref{fig:v_theta-N-F19-0200-0500-00000}(n)
 can be seen more obviously
 than those induced by 20-GeV WIMPs.
 In addition,
 comparing to Figs.~\ref{fig:v_theta-N-F19-0020-0500-00000}(k)
 to \ref{fig:v_theta-N-F19-0020-0500-00000}(n),
 the angular recoil--flux distribution
 in each corresponding energy window
 is now flatter.
 Interestingly,
 the inner--outer--sky asymmetry of
 the angular recoil--flux distribution
 in Fig.~\ref{fig:v_theta-N-F19-0200-0500-00000}(k)
 show a {\em reverse} pattern:
 (the decrease of) the recoil flux
 on the inner (east) sky
 is now clearly larger (and sharper)
 than that on the outer (west) sky.
 These ``WIMP--mass dependent'' characteristics indicate
 a possibility to pin down the mass of incident halo WIMPs
 by using the angular recoil--energy spectra
 (with different target nuclei)
 offered by directional direct detection experiments.

\subsection[200-GeV WIMPs off $\rmXe$ nuclei]
           {\boldmath
            200-GeV WIMPs off $\rmXe$ nuclei}
\label{sec:Xe129-0200}

 As in Sec.~\ref{sec:Xe129-0020},
 we consider the heavy nucleus $\rmXe$
 as our detector target.

 \def \Target       {Xe129}
 \InsertPlotvthetaN
  {200}
  {As Figs.~\ref{fig:v_theta-N-Xe129-0020-0500-00000}:
   the heavy nucleus $\rmXe$
   has been considered as our target,
   except that
   the mass of incident halo WIMPs is raised to $\mchi = 200$ GeV.%
   }

 First of all,
 comparing Figs.~\ref{fig:v_theta-N-Xe129-0200-0500-00000}(a)
 and \ref{fig:v_theta-N-Xe129-0200-0500-00000}(b)
 with Figs.~\ref{fig:v_theta-N-Xe129-0020-0500-00000}(a)
 and \ref{fig:v_theta-N-Xe129-0020-0500-00000}(b),
 sine $\alpha(\rmXe, \mchi = 200~{\rm GeV})$ is now pretty small ($\simeq 0.1$)
 and the nuclear form factor suppression becomes also much stronger
 \cite{DMDDD-fv_eff},
 the most frequent recoil angle
 would now be squeezed to only $\eta \simeq 75^{\circ}$.
 More precisely,
 in Figs.~\ref{fig:v_theta-N-Xe129-0200-0500-00000}(c)
 to \ref{fig:v_theta-N-Xe129-0200-0500-00000}(f)
 for each 5-keV energy window,
 the most frequent recoil angle
 would be at least $\eta \simeq 60^{\circ}$,
 or even as large as $\eta \simeq 80^{\circ}$.
 This means that
 a large number of observed recoil events
 would be deflected almost {\em perpendicularly}.
 Hence,
 the angular recoil--flux distributions 
 shown in Figs.~\ref{fig:v_theta-N-Xe129-0200-0500-00000}(g)
 to \ref{fig:v_theta-N-Xe129-0200-0500-00000}(j)
 would thus be extended much wider and flatter
 than those
 in Figs.~\ref{fig:v_theta-N-Xe129-0020-0500-00000}(g)
 to \ref{fig:v_theta-N-Xe129-0020-0500-00000}(j)
 (compare also
  the top frames of Figs.~23(a) and 23(c) of Ref.~\cite{DMDDD-NR}).

 Moreover,
 basically due to its heavy atomic mass
 and in turn
 the reduced transformation constant $\alpha$
 as well as
 the strong nuclear form factor suppression
 and thus the large recoil angle,
 with (heavy target nuclei like) $\rmXe$,
 the difference between
 the angular recoil--flux distributions
 in four considered energy windows
 shown in Figs.~\ref{fig:v_theta-N-Xe129-0020-0500-00000}
 and \ref{fig:v_theta-N-Xe129-0200-0500-00000}
 would now be strongly enlarged
 than those with (light target nuclei like) $\rmF$.
 One can also find that
 the differences between all corresponding plots
 in Figs.~\ref{fig:v_theta-N-F19-0200-0500-00000}
    and \ref{fig:v_theta-N-Xe129-0200-0500-00000}
 would be larger than the differences
 between plots
 in Figs.~\ref{fig:v_theta-N-F19-0020-0500-00000}
    and \ref{fig:v_theta-N-Xe129-0020-0500-00000}.
 Additionally,
 the ``reverse'' inner--outer--sky asymmetry of
 the angular recoil--flux distribution
 in the geocentric Galactic coordinate system
 observed in Fig.~\ref{fig:v_theta-N-F19-0200-0500-00000}(k)
 could also be found here,
 and,
 perhaps due to the heavy atomic mass of xenon nuclei,
 even in all four considered energy windows.
 These indicate clearly that,
 comparing and/or combining
 the angular recoil--energy spectra
 with different target nuclei
 could provide a method for reconstructing the WIMP mass
 analytically (and perhaps even model--independently).

%% file: sec-references.tex
%
%

%
%